\documentclass{jetpl}
\twocolumn
\title{Magnon BEC in superfluid $^3$He-A}

\rtitle{Magnon BEC in superfluid $^3$He-A}

\sodtitle{Magnon BEC in superfluid $^3$He-A}

\author{
Yu.M. Bunkov $^{\#}$\/\thanks{
yuriy.bunkov@grenoble.cnrs.fr; 
volovik@boojum.hut.fi} and
G.E. Volovik $^{*+}$
}

\rauthor{Yu.M. Bunkov, G.E.Volovik}

\sodauthor{Bunkov, Volovik}

\address{ $^{\#}$MCBT, Institut Neel, CNRS/UJF, Grenoble, 38042, France\\
$^{*}$ Low Temperature Laboratory, Helsinki University of
Technology, P.O.Box 5100, FIN-02015, HUT, Finland
\\
$^{+}$ Landau Institute for Theoretical Physics RAS, Kosygina 2,
119334 Moscow, Russia}

\dates{March 2, 2009}{*}

\abstract{The new mode of magnetization precession in superfluid $^3$He-A
in a squeezed aerogel has been
recently reported. We consider this mode in terms of the
Bose-Einstein condensation (BEC) of magnons. The difference
between magnon BEC states in $^3$He-A and in $^3$He-B is discussed.}

\begin{document}

\maketitle


The discovery \cite{HPD} and detailed investigations of the
phase-coherent precession of magnetization in superfluid $^3$He-B
generated a search for similar phenomena in other systems.
Superfluid $^3$He-A could be a proper system. However, it was found
that under typical conditions the coherent precession in $^3$He-A is
unstable \cite{InstabAB} because of the convex shape of spin-orbit
energy potential as function of magnetization
\cite{Fomin1979,Fomin1984}. It was suggested that the shape of
potential can be inverted and thus the coherent precession can be
stabilized if the orbital momentum of Cooper pairs in $^3$He-A is
oriented along the applied magnetic field \cite{BunkovVolovik1993}.
Recently such orientation has been reached for $^3$He-A immersed in
the axially squeezed aerogel \cite{Jap1}, and the first experiments
with the coherently precessing state (CPS) of magnetization have been
reported \cite{Sato2008}. Here we discuss the phenomenon of the
coherent precession of magnetization in superfluid $^3$He-A in terms
of the Bose-Einstein condensation (BEC) of magnons, and consider the
difference between magnon BEC states in $^3$He-A (CPS) and in
$^3$He-B (HPD).

BEC is one of the most remarkable
quantum phenomena. It corresponds to formation of collective quantum
state, in which the macroscopic number of particles is governed by a
single wave function. The formation of Bose-Einstein condensate was
predicted by Einstein in 1925, for review see e.g.
\cite{revBEC}. The almost perfect BEC state was observed in ultra
could atomic gases. In Bose liquids, the BEC is strongly modified by
interactions, but still remains the key mechanism for formation of
coherent quantum state, which experiences the phenomenon of superfluidity:
nondissipative superfluid current. In liquid $^4$He the depletion of
the condensate is very large: in the limit of zero temperature only
about 10\%  of particles occupy the state with zero momentum.
Nevertheless the whole liquid (100\% of atoms) forms the coherent
quantum state at $T=0$, so that the superfluid density
equals the total density,
$\rho_s(T=0)=\rho$. The latter is valid for  any
monoatomic superfluid system with translational invariance,
including superfluid $^3$He with the non BEC mechanism of coherent
quantum state. If  translational invariance is violated by
impurities, crystal fields or other inhomogeneity, the superfluid
component  is suppressed: $\rho_s(T=0)<\rho$.

Superfluidity is a very general quantum property of  matter at low
temperatures, with variety of possible nondissipative superfluid
currents, such as  supercurrent of electric charge in superconductors;
hypercharge supercurrent in the vacuum of Standard Model;
supercurrent of color charge in a dense quark matter; etc.
The origin of superfluidity is the spontaneous violation of the  $U(1)$
symmetry related to the conservation of the corresponding charge or particle
number. That is why, strictly
speaking, the theory of superfluidity is applicable to systems
with conserved  charge  or particle number. However, it can be extended to
systems with a weakly violated conservation law. This means that a
system of sufficiently  long-lived quasiparticles, such as phonons,
rotons, spin waves  (magnons), excitons,  etc.,  can also form the coherent
state, which is close to thermodynamic equilibrium state of Bose condensate.

The phase-coherent precession of magnetization in superfluid
$^3$He-B discovered in 1984 \cite{HPD} can be considered as a
realization of superfluidity of quasiparticles, which  results from
the BEC of  magnon quasiparticles \cite{BunkovVolovik2008,v}. 
In $^3$He-B, the
magnon BEC is represented by a domain with a fully phase-coherent
precession of magnetization, known as the Homogeneously Precessing
Domain (HPD). HPD exhibits all the properties of spin superfluidity
(see Reviews \cite{v,FominLT19,BunkovHPDReview}). These include in
particular: spin supercurrent which transports the magnetization;
spin current Josephson effect and phase-slip processes at the
critical current; and spin current vortex  --  a topological defect
which is the analog of a quantized vortex in superfluids and of an
Abrikosov vortex in superconductors; etc. The temperature at which
the BEC in $^3$He-B exists is by several orders of magnitude smaller
than the transition temperature of magnon Bose condensation  \cite{v}. This
implies that the gas of magnons forms practically 100\% BEC state,
even if only  a small number  of excitations is originally pumped
into the $^3$He-B sample. Magnon condensation to the lowest energy
states has been also found in yttrium-iron garnet
\cite{Demokritov,Demidov} with the fraction of the condensed magnons
being less than 1$\%$.  Condensation of such quasiparticles as
polaritons to the lowest energy states has been reported in Refs.
\cite{Kasprzak,Kasprzak2008}. The polariton condensate is formed as
dynamical out-of-equilibrium state, which is rather far  from the
true thermodynamic BEC.

 In bulk  $^3$He-A, the BEC of magnons is unstable because
of attractive interaction between magnons, which is reflected in the
concave shape of the spin-orbit (dipole-dipole) interaction
potential. However, under special conditions, when $^3$He-A is
confined in the properly deformed aerogel, interaction between
magnons  becomes repulsive and a stable Bose condensate is formed.
The magnon BEC is described in terms of complex order parameter
$\Psi$, which is related to the precessing spin in the following way
\cite{v,FominLT19}:
\begin{eqnarray}
  \Psi=\sqrt{2S/\hbar}~\sin \frac{\beta}{2}~e^{i\omega t+i\alpha}~,~
  \\
S_x+iS_y =S\sin\beta ~e^{i\omega t+i\alpha}, \label{OrderParameter}\\
 N_M=\int d^3r \vert \Psi\vert^2 =\int d^3r \frac{S-S_z}{\hbar}~.
\label{ChargeQ}
\end{eqnarray}
Here ${\bf S}=(S_x,S_y,S_z=S\cos\beta)$ is the vector of spin
density; $\beta$ is the tipping angle of precessing magnetization;
$\omega$ is the precession frequency (in the regime of continuous
NMR, it is the frequency $\omega_{\rm RF}$ of the applied RF field
and it plays the role of  the chemical potential $\mu=\omega$ for
magnons); $\alpha$ is the phase of precession; $S$ is  the
equilibrium value of spin density in the applied magnetic  field
${\bf H}=H\hat{\bf z}$ (in $^3$He liquids $S=\chi H/\gamma$, where
$\chi$ is spin suscepti\-bility of $^3$He-B or $^3$He-A, and
$\gamma$ the gyromagnetic ratio of the $^3$He atom);
$\vert\Psi\vert^2=n_M$ is the density of magnons and $N_M$ is the
total   number of magnons in the precessing state.

The corresponding Gross-Pitaevskii equation is
  \begin{eqnarray}
 \frac{\delta F}{\delta \Psi^*}=0~,
  \label{GP}
  \\
 F=\int d^3r\left(\frac{\vert\nabla\Psi\vert^2}{2m_M} +(\omega_L({\bf r})-\omega)\vert\Psi\vert^2+F_D\right).
\label{GL}
\end{eqnarray}
Here  $\omega_L({\bf r})=\gamma H({\bf r})$  is the
local Larmor frequency which plays the role of external potential
for magnons;   $m_M$ is the
magnon mass; and $F_D$ is  the  spin-orbit  interaction
averaged over the fast precession, which plays the part of
interaction between magnons. In superfluid $^3$He, $F_D$ depends
on interaction between spin
and orbital degrees of freedom and is
determined by the direction $\hat{\bf l}$ of the orbital angular
momentum  of Cooper pairs. For $^3$He-A, it has the form \cite{BunkovVolovik1993}
\begin{eqnarray}
  F_D= \frac {\chi\Omega_L^2}{4} \times
    {\nonumber}
  \\
  \left[ -2\frac{\vert\Psi\vert^2}{S} +
  \frac{\vert\Psi\vert^4}{S^2}    +
    \left( -2+4 \frac{\vert\Psi\vert^2}{S}  -
  \frac {7}{4}\frac{\vert\Psi\vert^4}{S^2}\right)\sin^2{\beta_L}\right],
  \label{FD}
  \end{eqnarray}
where $\beta_L$ is the angle of $\hat{\bf l}$ with respect to
magnetic field; and $\Omega_L\ll \omega_L$ is the Leggett frequency
characterizing the spin-orbit coupling (we put  $\hbar=\gamma=1$).

While the  sign of the  quadratic term in Eq.(\ref{FD})  is not
important because it only leads to the shift of the chemical
potential $\mu\equiv \omega$ in Eq.(\ref{GL}), the sign of quartic
term is crucial for stability of BEC. In a static bulk $^3$He-A,
when $\Psi = 0$,  the spin-orbit energy $F_D$ in  Eq.(\ref{FD}) is minimized when $\hat{\bf l}$
is perpendicular to magnetic field, $\sin\beta_L = 1$. Then one has
\begin{equation}
  F_D= \frac {\chi\Omega_L^2}{4}
  \left[-2+ 2\frac{\vert\Psi\vert^2}{S} -  \frac {3}{4}
  \frac{\vert\Psi\vert^4}{S^2}    \right],
  \label{FDbeta_L90}
  \end{equation}
with the negative quartic term. The attractive interaction between
magnons destabilizes the BEC, which means that homogeneous
precession of magnetization in $^3$He-A becomes unstable, as was
predicted by Fomin \cite{Fomin1979} and observed experimentally in
Kapitza Institute \cite{InstabAB}.

However, as follows from \eqref{FD}, at sufficiently large magnon
density $n_M=\vert\Psi\vert^2$
 \begin{equation}
 \frac{8+\sqrt{8}}{7}~S> n_M> \frac{8-\sqrt{8}}{7}~S~,
\label{PositiveCondition}
\end{equation}
the factor in front of $\sin^2\beta_L$ becomes positive. Therefore
it becomes energetically favorable to  orient the orbital momentum
along the magnetic field, $\beta_L=0$, and after that the quartic
term in Eq.(\ref{FD})  becomes positive. In other words, with increasing the density
of Bose condensate, the originally attractive interaction between
bosons should spontaneously become repulsive when the critical
magnon density $ n_M= S(8-\sqrt{8})/7$ is reached. If this happens, the magnon BEC becomes
stable and in this way the state with coherent precession (CPS)
could be formed \cite{BunkovVolovik1993}. This self-stabilization
effect is similar to the effect of $Q$-ball, where bosons create the
potential well in which they condense (on theory and experiment of
magnon condensation into $Q$-ball  see Ref.
\cite{BunkovVolovik2007}). However, such a self-sustaining BEC with
originally attractive boson interaction has not been achieved
experimentally in bulk $^3$He-A, most probably because of the large dissipation, due
to which the threshold value of the condensate density has not been
reached. Finally the fixed orientation of $\hat{\bf l}$ has been
achieved in $^3$He-A confined in aerogel -- the material with high
porosity, which is about 98\%
 of volume in our experiments. As a consequence,  when magnetic field
 was oriented along $\hat{\bf l}$ (i.e. in geometry with $\beta_L=0$), the
 first indication of coherent precession  in $^3$He-A has been reported \cite{Sato2008}.

Silicon strands of aerogel play the role of impurities with local
anisotropy along the strands.  According to the Larkin-Imry-Ma effect, the
random anisotropy  suppresses the orientational long-range order of the orbital vector $\hat{\bf l}$; however, when the aerogel sample is deformed the
long-range order of $\hat{\bf l}$ is restored
\cite{VolovikAerogel}.  Experiments with globally squeezed aerogel
\cite{Jap1} demonstrated that a uni-axial deformation by about 1\%
is sufficient for global orientation of the vector $\hat{\bf l}$
along ${\bf H}$: the observed shape of the NMR line has a large
negative frequency shift  corresponding to $\beta_L$ spreading from
0 to about 20$^\circ$. Previously the small negative frequency shift
has been observed in some samples of aerogel \cite{Osh,
Dmitriev2006}, which can be explained by a residual deformations of
these samples.

Let us first consider a perfect aerogel sample with global
orientation of the vector $\hat{\bf l}$ along ${\bf H}$.
 For $\beta_L=0$, the GL free energy acquires the standard form:
\begin{equation}
 F=\int d^3r\left(\frac{\vert\nabla\Psi\vert^2}{2m_M} +(\omega_L({\bf r})-\mu)\vert\Psi\vert^2+\frac{1}{2}b\vert\Psi\vert^4\right),
\label{GL2}
\end{equation}
where we modified the chemical potential by the constant frequency shift:
\begin{equation}
\mu= \omega + \frac{\Omega_L^2}{2\omega} ~,
   \label{TildeOmega}
   \end{equation}
   and the parameter $b$ of repulsive magnon interaction is
 \begin{equation}
b= \frac{\Omega_L^2}{2\omega S}
 \label{b}
  \end{equation}
At $\mu>\omega_L$, magnon BEC must be formed with
density
\begin{equation}
\vert\Psi\vert^2=\frac{\mu-\omega_L}{b} ~.
   \label{EquilibriunPsi}
\end{equation}
This is distinct from $^3$He-B, where condensation starts with finite condensate density.
Eq. (\ref{EquilibriunPsi}) corresponds to the following dependence of  the frequency
shift on tipping angle $\beta$ of coherence precession:
 \begin{equation}
\omega - \omega_L= - \frac{\Omega_L^2}{2\omega} \cos\beta~.
 \label{freqshiftNoRF}
\end{equation}
If the precession is induced by continuous wave (CW) NMR, one should also
add the interaction with the RF field, ${\bf H}_{\rm RF}$, which is
transverse to the applied constant field ${\bf H}$.  In CW NMR experiments,
the RF field prescribes the frequency of
precession, $\omega=\omega_{\rm RF}$, and thus fixes the chemical
potential $\mu$. In the precession frame, where both the RF field
and the spin density ${\bf S}$ are constant in time,  the interaction term is
\begin{equation}
F_{\rm RF}=-\gamma {\bf H}_{\rm RF}\cdot{\bf S}= -  \gamma H_{\rm
RF} S_\perp \cos(\alpha-\alpha_{\rm RF}) ~, \label{InteractionRF}
\end{equation}
where $H_{\rm RF}$ and $\alpha_{\rm RF}$ are the amplitude and the phase of
the RF field. In the language of magnon BEC, this term  softly
breaks the $U(1)$-symmetry and serves as a source of magnons
\cite{Volovik2008}:
\begin{equation}
F_{\rm RF}(\psi)= - \frac{1}{2}\eta\left(\psi +\psi^*\right) ~,
\label{SymmetryBreakingRF1}
\end{equation}
which compensates the loss of magnons due to magnetic relaxation.
The symmetry-breaking field $\eta$ is:
\begin{equation}
\eta=  \gamma H_{\rm RF} \sqrt{2S - n_M}  ~.
\label{SymmetryBreakingField2}
\end{equation}
The phase difference  between the condensate and the RF field, $\alpha-\alpha_{RF}$,
 is determined by the
energy losses due to magnetic relaxation, which is compensated by
the pumping of power from the CW RF field:
\begin{equation}
  W_+ = \omega SH_{\rm RF}\sin\beta \sin{(\alpha-\alpha_{\rm RF})}~.
  \label{pumping}
\end{equation}
The phase shift is automatically adjusted to compensate the losses. If dissipation is
small, the phase shift is small,  $\alpha-\alpha_{\rm RF}\ll 1$, and can be neglected.
The neglected quadratic  term $(\alpha-\alpha_{\rm RF})^2$ leads to the nonzero mass of the
Goldstone boson -- quantum of sound waves (phonon)
in the magnon superfluid \cite{Volovik2008}; in $^3$He-B
the phonon mass, which is proportional to $\sqrt{H_{\rm RF}}$, has been measured
\cite{PhononMass,Skyba}.
In the limit of small dissipation the main role of the RF field is to modify the profile
of the GL free energy in \eqref{GL2} by adding the term:
 \begin{equation}
F_{\rm RF}(n_M)= - \eta|\psi|= - \gamma H_{\rm
RF}\sqrt{n_M(2S-n_M)}~. \label{SymmetryBreakingRF3}
\end{equation}
 Equation $dF/dn_M=0$ now gives the following modification  of Eq.\eqref{freqshiftNoRF}
 for NMR frequency shift  as function of magnon density $n_M= \vert\Psi\vert^2=S(1-\cos\beta)$:
 \begin{equation}
\omega - \omega_L= - \frac{\Omega_L^2}{2\omega_L} \cos\beta - \gamma
H_{\rm RF} \cot \beta.
 \label{freqshift}
\end{equation}
 For finite
$\alpha-\alpha_{\rm RF}$, the last term should be multiplied by
$\cos(\alpha-\alpha_{\rm RF})$. According to Ref. \cite{Sato2008},
the energy losses are proportional to square of transverse
magnetization, and thus we have:
\begin{equation}
  W_- = \sigma \sin^2\beta~,
  \label{losses}
\end{equation}
where $\sigma$ is the phenomenological parameter.

\begin{figure}[ttt]
\includegraphics[width=0.5\textwidth]{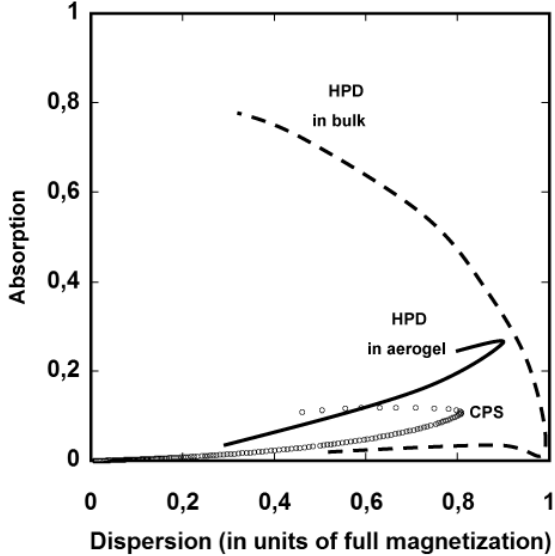}
\caption{Fig. 1. Typical absorption/dispersion relation for different states of coherent precession.
HPD in the bulk $^3$He-B is shown schematically. Signals from  HPD in  $^3$He-B in aerogel and from the CPS state in  $^3$He-A in aerogel
correspond to experimental data in Ref. \cite{Sato2008}. }
 \label{absorption/dispersion}
\end{figure}

Since the pumping \eqref{pumping} is proportional to
$\sin\beta\sin(\alpha-\alpha_{\rm RF})$, then there must be a
critical tipping angle $\beta_c$, at which the pumping cannot
compensate the losses, $\sin \beta_c =  \omega S H_{RF} /\sigma$,
and the coherent precession collapses. For a homogeneous case the
critical angle should corresponds to $\alpha_c = \alpha-\alpha_{\rm
RF} = 90^\circ$. For the real case, it is instructive to consider the phase portrait of
the CW NMR signal measured in experiments:  
the time development of the signal in the plane of 
absorption $M_\perp \sin{(\alpha-\alpha_{\rm RF})}$
and dispersion $M_\perp \cos{(\alpha-\alpha_{\rm RF})}$, where $M_\perp$ is the
total transverse magnetization in the cell. 
This diagram demonstrates the time dependence
of angle $\alpha-\alpha_{\rm RF}$ during sweeping the frequency shift
$\omega-\omega_L$ (actually the field $H$ is swept).
 For the HPD state in bulk $^3$He-B this is shown by dashed line in Fig.1. 
At first stage the precessing domain is filling
the cell. During the process of filling the absorption remains rather small,
and thus dispersion corresponds to the full transverse magnetization 
$M_\perp$ which grows linearly with growing domain. 
After domain fills the whole cell, the full 
transverse magnetization is fixed, and the signal follows  the circle around the origin. 
This correspond to increase of angle $\alpha-\alpha_{\rm RF}$ due to increasing of relaxation. 
Finally the coherent precession collapses. The critical angle $\alpha_c$ for HPD is
typically about $70^\circ - 45^\circ$. It is significantly smaller
than expected $90^\circ$, and it is due to inhomogeneity of relaxation. In the region of
larger magnetic (Leggett-Takagi) relaxation, the local $\alpha({\bf r})-\alpha_{\rm
RF}$ is larger. The spatial gradient of angle $\alpha$ generates the
spin supercurrent, which transports the magnetization and supports the
coherence of precession. The HPD state collapses, when the local angle of
precession reaches the $90^\circ$ in some region of the cell.

\begin{figure}[ttt]
\includegraphics[width=0.53\textwidth]{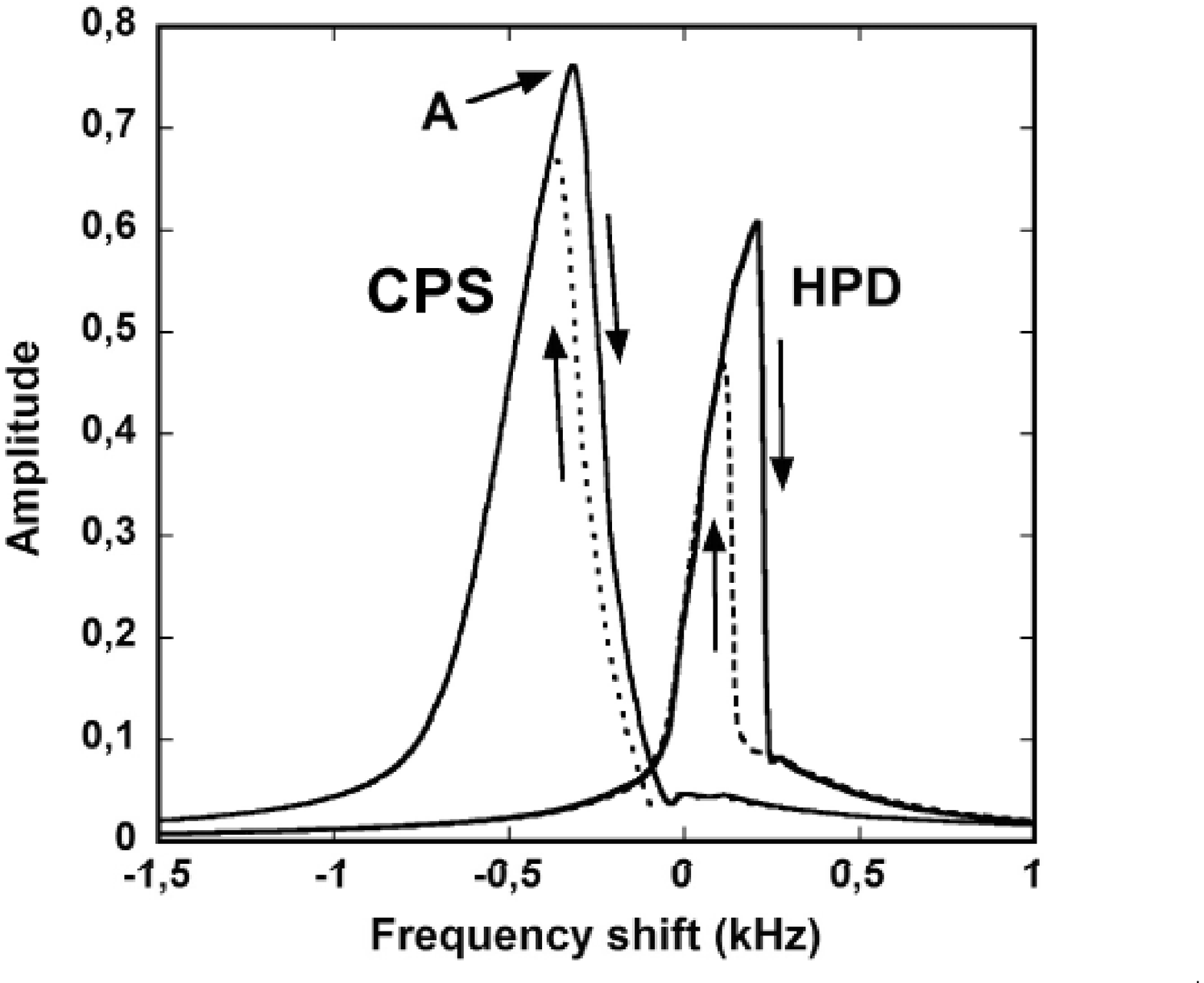}
\caption{Fig. 2. Formation of the homogeneously precessing domain (HPD) in
$^3$He-B  and of phase-coherent precession (CPS) in $^3$He-A  under
upward and downward frequency sweeps. The experimental data are from
\cite{Sato2008}. }
 \label{formation}
\end{figure}

A similar behavior was found both for HPD and CPS states in aerogel,
as shown in Fig.1. But there is also a peculiar difference which
is related to a spatial inhomogeneity of the aerogel sample. 
The first HPD in $^3$He-B in aerogel was observed on a very inhomogeneous
sample. As a result, the HPD was created locally and was not
able to grow through the whole sample \cite{BunkovaeroHPD}. Later on this
experiment was repeated with the more homogeneous aerogel
\cite{DmitHPD}, and HPD  filled the whole sample. Indeed, the
aerogel samples of such quality are very rear.

The sample of aerogel, used in the work \cite{Sato2008}, can be
considered as of intermediated quality. The HPD signal can be created
in the whole cell, but the absorption signal is rather big. The
spatial inhomogeneity of absorption is also clear from a small value
of threshold  $\alpha_c$, at which collapse occurs: it is about $15^\circ$ in
 Fig. 1. The threshold $\sin \beta_c$ was found to increase
with increasing $H_{\rm RF}$ (see Fig.3 in Ref. \cite{Sato2008}), and
the precession with large $\beta$ was finally achieved at large $H_{\rm RF}$. 
At small
$\beta$ the influence of the diverging $ H_{\rm RF} \cot \beta$ term in
\eqref{freqshift} is clearly seen, while at large $\beta$,  signals
at different excitations fall onto a universal curve independent of
the amplitude of the RF-field, as shown in the Fig. 3 of Ref.
\cite{Sato2008}. This demonstrates that at large $\beta$ the magnon
BEC is self-consistent and is not sensitive to the RF-field; the
latter is only needed for compensation of the spin and energy losses.
The amplitude of highest RF field, used in the experiments of
Ref.\cite{Sato2008} was only $0.05$ Oe, which is much smaller than the
frequency shift and the inhomogeneity of the NMR line.

What can be origin of the spatial  inhomogeneity of relaxation? The candidates
for regions with high dissipation could be topological defects, such
as solitons -- domain walls between two possible orientations of
vector $\hat{\bf l}$ in the deformed aerogel: parallel and
anti-parallel to ${\bf H}$. If the density of solitons is relatively
small, and thus the regions with small dissipation are dominating,
the observed average value $\left<\sin(\alpha({\bf r})-\alpha_{\rm
RF})\right>$ will be small. In Fig. 2 we reproduce the signals from
Ref.\cite{Sato2008}. The important difference between the CPS and
HPD signals, observed in aerogel, and the HPD signal in bulk is the fact
that after collapse the states in aerogel are not
destroyed completely: the coherent precession survives in some parts of the
sample. Furthermore, the state can be excited by sweeping the frequency
back. The latter shows, that there are some regions in the sample with
a very different orientation of the order parameter, which can get the
energy from the RF field, transport  magnetization
by spin supercurrents to the other parts of the cell and restore the CPS
in the whole cell. This is a  natural explanation, whose justification
 however requires the detailed knowledge
of  the distribution of inhomogeneity and texture of the order parameter in aerogel.

Of course, the final proof of the coherence of precession in $^3$He-A in aerogel
would be the observation of the free precession after a pulsed NMR or after a switch
off the CW NMR. However, it is not excluded that what was observed
in Ref.\cite{Sato2008} corresponds not to a single domain of precession, but
to a few weakly interacting CPS domains, which are kept in phase by the RF field rather than by supercurrents between them. In this case it would not be so easy to detect the coherent precession
in the pulsed NMR. Example of such kind is provided by  experiments with
 bulk $^3$He-B,  which was divided into 5 independent parts by
maylar foils \cite{HPDFull}. Then in the CW NMR the signal corresponded
to all 5 HPD states being in phase, while in pulsed NMR the coherence between the HPD state was lost
and beating of 5
independent HPD states was observed.

In conclusion, in contrast to the homogeneously precessing domain
(HPD) in $^3$He-B,  the magnon Bose condensation in $^3$He-A obeys
the standard Gross-Pitaevskii equation. In bulk $^3$He-A, the Bose
condensate of magnons is unstable because of the attractive
interaction between magnons. In $^3$He-A confined in aerogel, the
repulsive interaction is achieved by the proper deformation of the
aerogel sample, and the Bose condensate becomes stable. The magnon
BEC in $^3$He-A adds to the other two coherent states of magnons
observed in $^3$He-B:  HPD state   and $Q$-ball
\cite{BunkovVolovik2007}. New experiments to observe the superfluid
phenomena accompanying the coherent precession in $^3$He-A  (spin
supercurrent transport of magnetization, Josephson phenomena and
spin-current vortices) are expected in future. It would be
interesting to search for similar dynamical coherent states of
excitations in other condensed matter systems (see for example Refs.
\cite{Demokritov,Demidov,Kasprzak,Kasprzak2008}).

 \section*{\hspace*{-4.5mm}ACKNOWLEDGMENTS}
\noindent It is a pleasure to thank T. Mizusaki, M. Kubota,
A. Matsubara for numerous discussions and V. Dmitriev
for helpful criticism. This work was supported by the program ULTI of the
European Union (contract number: RITA-CT-2003-505313) and the
project of collaboration between  of CNRS and Russian Academy of
science No 21253. GEV is supported in part by the Russian Foundation
for Basic Research (grant 06--02--16002--a) and the
Khalatnikov--Starobinsky leading scientific school (grant
4899.2008.2)


\end{document}